\documentclass[article]{jss_arxiv}
\usepackage{amssymb}

\author{Giri Gopalan\\Harvard University \And 
        Luke Bornn\\Harvard University}
\title{\pkg{FastGP}: an R package for Gaussian processes}

\Plainauthor{Giri Gopalan, Luke Bornn} 
\Plaintitle{\pkg{FastGP}: an R package for Gaussian processes and multivariate normal distributions}
\Shorttitle{\pkg{FastGP}: an R package for Gaussian processes} 

\Abstract{
Despite their promise and ubiquity, Gaussian processes (GPs) can be difficult to use in practice due to the computational impediments of fitting and sampling from them. Here we discuss a short R package for efficient multivariate normal functions which uses the \pkg{Rcpp} and \pkg{RcppEigen} packages at its core. GPs have properties that allow standard functions to be sped up; as an example we include functionality for Toeplitz matrices whose inverse can be computed in $O(n^2)$ time with methods due to Trench and Durbin \citep{gol96}, which is particularly apt when time points (or spatial locations) of a Gaussian process are evenly spaced, since the associated covariance matrix is Toeplitz in this case. Additionally, we include functionality to sample from a latent variable Gaussian process model with elliptical slice sampling \citep{mur10}. 
}
\Plainkeywords{Gaussian processes, multivariate normal distributions, Rcpp, RcppEigen} 

\Address{
  Giri Gopalan\\
  Department of Statistics\\
  Harvard University\\
  1 Oxford Street\\
  E-mail: \email{giri@alumni.caltech.edu}
}

\begin{document}


\section[Introduction]{Introduction}
Many methodologies involving a Gaussian process rely heavily on computationally expensive functions such as matrix inversion and the Cholesky decomposition. Rather than create a package to solve a particular high-level Gaussian process (GP) task (e.g., expectation propagation, variational inference, regression and classification \citep{neal98}), the aim of \pkg{FastGP} is to improve the performance of these fundamental functions in order to help all researchers working with GPs. While there exist R packages for sampling from a  multivariate normal  distribution (MVN) or evaluating the density of an MVN, notably \pkg{MASS} and \pkg{mvtnorm} on CRAN \citep{gen14,ven02}, we have found such packages can be slow in the context of GPs, partially due to unnecessary checks for symmetry and positive definiteness (which hold for GPs with commonly used kernels such as squared exponential or Matern \citep{ras05}) or not accounting for the structure (e.g. Toeplitz) of the underlying covariance matrix. Hence, we write functions optimized with \pkg{Rcpp} and \pkg{RcppEigen} \citep{bat13, edd13} to make these tasks more computationally efficient, and demonstrate their efficiency by benchmarking them against built-in R functions and methods from the \pkg{MASS} and \pkg{mvtnorm} libraries. Additionally, we include functionality to sample from the posterior of a Bayesian model for which the prior distribution is multivariate normal using elliptical slice sampling, a task which is often used alongside GPs and due to its iterative nature, benefits from a C++ version \citep{mur10}.
\\
\\
To elaborate, a Gaussian process (GP) is a collection of random variables (i.e., a stochastic process) $(X_t)$ such that any finite subset of these random variables has a joint multivariate normal distribution \citep{gri01}. Such processes have been used extensively in recent decades, particularly in machine learning, spatial and temporal statistics, and computer experiments \citep{ras05}. However, GPs can be difficult to work with in practice because they are computationally onerous; to be precise, the density of a multivariate normal  (MVN) vector in $\mathbb{R}^n$ is:
\begin{eqnarray}
f(x) &=& \frac{1}{(2\pi)^{n/2}|\Sigma|^{1/2}}\exp^{-(x-\mu)^T\Sigma^{-1}(x-\mu)/2}
\end{eqnarray}
This involves the computation of a determinant and inverse of a matrix in $\mathbb{R}^{n \times n}$, generally taking $O(n^3)$ operations to complete, and the computation of the Cholesky decomposition is typically a prerequisite for sampling from a multivariate normal, which also takes $O(n^3)$ time to complete \citep{gol96}. Many spatial models, including those which tackle nonstationarity as in \citet{bornn}, parametrize the covariance matrix $\Sigma$,  and hence for Monte Carlo-based inference require repeated recalculations of $\Sigma^{-1}$ and $|\Sigma|$.
\section[Key Functions]{Key functions and benchmark results} 
\subsection{Key functions and package organization}
The core functions of \pkg{FastGP} can be categorized into three sets. The first set of functions are matrix operations that are necessary for sampling from and evaluating the density of a multivariate normal random variable. These are: a function for inversion using \pkg{RcppEigen}, a function for inverting a symmetric positive definite Toeplitz matrix  in $O(n^2)$ time (which, as aforementioned, can be useful for inverting a covariance matrix in which the underlying points are evenly spaced \citep{sto99}) which uses methods due to Trench and Durbin \citep{gol96} written in \pkg{Rcpp}, and a function for evaluating the Cholesky decomposition of a matrix using \pkg{RcppEigen}. To be as explicit as possible, the inversion and Cholesky decomposition come directly from \pkg{RcppEigen} \citep{bat13}. The second set of functions are those which directly simulate from and evaluate the log density of a multivariate normal, both in the general case and when the underlying covariance matrix is Toeplitz. The final major function included in the package is the elliptical slice sampling algorithm for simulating from the posterior of a Bayesian model in which the prior is jointly multivariate normal, which tends to outperform classical methods such as Metropolis-Hastings computationally, as is evidenced empirically by \citet{mur10}. 

\subsection{Benchmark results}
Here we use the \pkg{rbenchmark} \citep{eug08} package to demonstrate the efficacy of these methods. In particular we test these functions with a mock covariance matrix from a square exponential Gaussian process on 200 evenly spaced time points with $\sigma$ and $\phi$ arbitrarily set to 1 (and hence the covariance matrix is Toeplitz in this case). 

\begin{table}[ht]
\caption{Benchmarking the  runtime for functions included in \pkg{FastGP}. The numbers indicate how many times faster the functions performed using our package versus standard R, \pkg{MASS}, and \pkg{mvtnorm} functions, using the \texttt{benchmark} function from the \pkg{rbenchmark} package.}
\centering
\begin{tabular}{c c c c}
\hline\hline
\pkg{FastGP} Function & Standard Function  & Relative Speed Improvement\\ [0.5ex] 
\texttt{rcppeigen\_invert\_matrix} & \texttt{solve} & x3.287 \\
\texttt{tinv} & \texttt{solve} & x14.374 \\ 
\texttt{rcppeigen\_get\_det} & \texttt{det} & x1.851 \\
\texttt{rcpp\_log\_dmvnorm, istoep= TRUE} & \texttt{dmvnorm} & x1.966 \\
\texttt{rcpp\_log\_dmvnorm, istoep= FALSE} & \texttt{dmvnorm} & x.6592 \\
\texttt{rcpp\_rmvnorm}  &  \texttt{rmvnorm} & x23.462 \\
\texttt{rcpp\_rmvnorm}  &  \texttt{mvrnorm} & x22.812 \\ 
\hline

\hline
\end{tabular}
\label{table:nonlin}
\end{table}

\subsection{Demonstration of elliptical slice sampling}
We consider a model where we observe a ``warped" signal $s =A\sin((t+w)/T)+\epsilon$ where $w$ is drawn according to a 0 mean GP with squared exponential kernel, $\epsilon$ is drawn according to a normal distribution with 0 mean and $\sigma = .001$, and $A$ and $T$ are known constants. Our objective is to perform inference on the latent warping $w$, and we can do this with the elliptical slice sampling function included with \pkg{FastGP}, as in \texttt{FastGPdemo.r}. The function implements the algorithm as described by \citet{mur10} and benefits from the use of the optimized functions contained within \pkg{FastGP} since each iteration requires several log-likelihood evaluations (which may require the evaluation of the log-density of a multivariate normal distribution) and drawing from a multivariate normal distribution. 
This results in the following posterior draws for $w$ illustrated in \textbf{Figure 1}. Additionally we benchmark  an elliptical slice sampler with \pkg{FastGP} functions versus an elliptical slice sampler with standard functions below.
\begin{table}[ht]
\caption{Benchmarking the runtime for elliptical slice sampling using functions from \pkg{FastGP} versus elliptical slice sampling using standard functions from R, \pkg{mvtnorm}, and \pkg{MASS}.}
\centering
\begin{tabular}{c c c c}
\hline\hline
\pkg{FastGP} Function & Standard Function  & Relative Speed Improvement\\ [0.5ex] 
\texttt{rcpp\_ess} & \texttt{standard\_ess} & x1.508 \\
\hline

\hline
\end{tabular}
\label{table:nonlin}
\end{table}
\begin{figure}
\vspace{-10ex}
\includegraphics[width=1.0\linewidth]{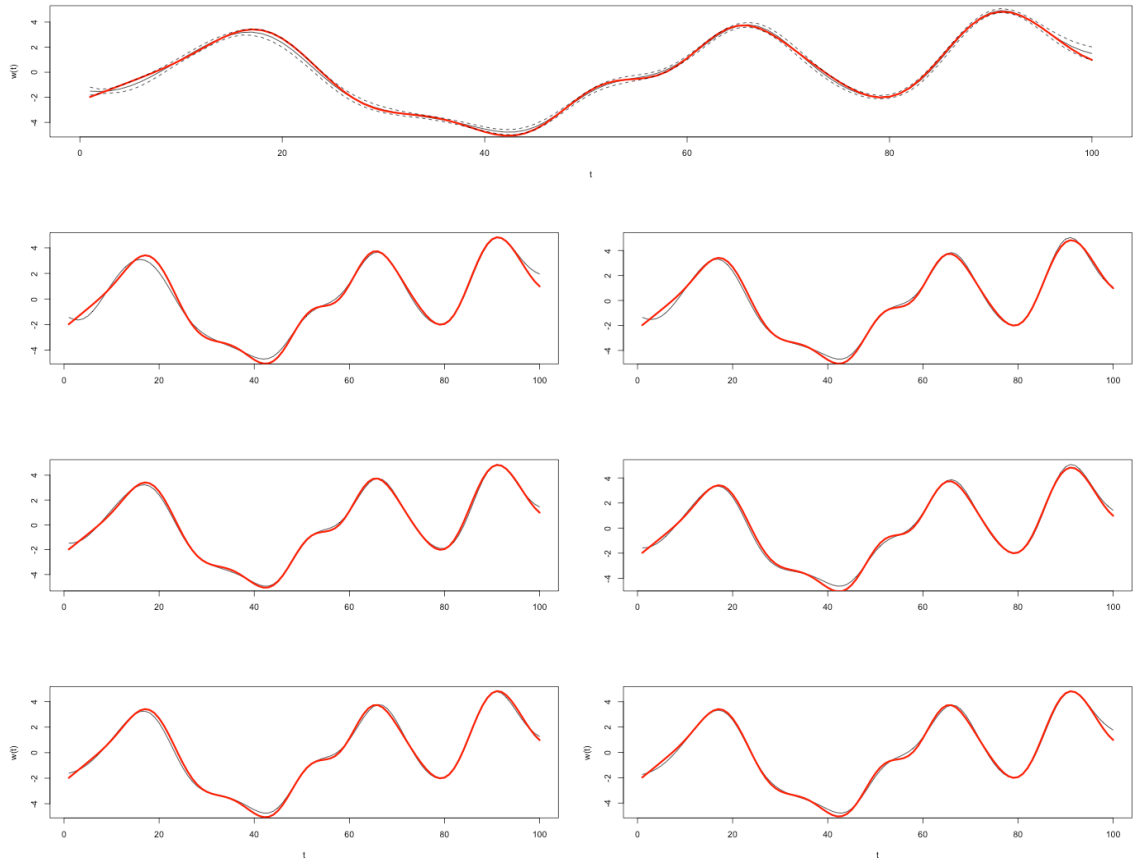}
\caption{Top: The red line indicates ground truth, the black line indicates the mean of the posterior latent samples, and the dotted lines indicate the 2 standard deviations above and below the mean, respectively, of the latent samples. Remaining rows show the 100th, 300th, 500th, 700th, 900th, and 1000th  MCMC samples for the warping $w$ respectively, in black, compared to ground truth in red.}
\end{figure}
\section[Conclusion]{Conclusion}
To summarize, we have written an R package \pkg{FastGP} using \pkg{Rcpp} and \pkg{RcppEigen} for handling multivariate normal distributions in the context of Gaussian processes efficiently. Additionally we have included functionality to perform Bayesian inference for latent variable Gaussian process models with elliptical slice sampling \citep{mur10}.

\end{document}